# LARGE-SCALE DATA MODELLING IN HIVE AND DISTRIBUTED QUERY PROCESSING USING MAPREDUCE AND TEZ

Abzetdin Adamov

*Center for Data Analytics Research (CeDAR)*
*ADA University, Baku, Azerbaijan*
*aadamov@ada.edu.az*

**Abstract**

*Huge amounts of data being generated continuously by digitally interconnected systems of humans, organizations and machines. Data comes in variety of formats including structured, unstructured and semi-structured, what makes it impossible to apply the same standard approaches, techniques and algorithms to manage and process this data. Fortunately, the enterprise level distributed platform named Hadoop Ecosystem exists.*

*This paper explores Apache Hive component that provides full stack data managements functionality in terms of Data Definition, Data Manipulation and Data Processing. Hive is a data warehouse system, which works with structured data stored in tables. Since, Hive works on top the Hadoop HDSFS, it benefits from extraordinary feature of HDFS including Fault Tolerance, Reliability, High Availability, Scalability, etc. In addition, Hive can take advantage of distributed computing power of the cluster through assigning jobs to MapReduce, Tez and Spark engines to run complex queries. The paper is focused on studying of Hive Data Model and analysis of processing performance done by MapReduce and Tez.*

# INTRODUCTION

Hadoop is open source common platform that combines two main tasks of any operating system: storing and processing data. Unlike to traditional systems, Hadoop accomplishes that tasks towards Big Data. The popularity of Hadoop increases day by day, because of simplicity, scalability and affordability that it enables thanks to its distributed architecture. Although, the Hadoop Core consists of two main components (HDFS and MapReduce) and has limited functionality, but thanks to many other components available in the Hadoop Ecosystem under Apache Licence, this platform can cover any requirements to manage and process data regardless to its size and format.

Data Analytics in scale that is enabled by Hadoop Ecosystem opens new horizons in turning operational data of businesses into actionable knowledge and consequently into value. In most cases, operational data is generated in structured format and stored in RDBMS. These databases and warehouses are still important, but now in era of Big Data businesses deal with amounts of data that can't feet into traditional RDBMS. Another challenge here is that in order to keep competitive advantage, businesses want to use for analytics all available data including website clickstream data, text from call centers, emails, instant messaging repositories, open data initiatives from public and private entities. Its clear that this goal can't be achieved based on traditional RDBMS systems. In contrast, Hadoop is a platform which consists of many components designed to accomplish specific tasks using particular data format. Hadoop Ecosystem components are classified into several categories to make it easier for user to choose appropriate components in accordance to the functions they designed for. There are following categories of the Hadoop components:

- Core Hadoop
- Governance, Integration
- Data Access and Storage
- Operations, Monitoring, Orchestration
- Security
- Data Intelligence

This paper explores Apache Hive component that provides full stack data managements functionality in terms of Data Definition, Data Manipulation and Data Processing. Hive is a data warehouse system, which works with structured data stored in tables. Since, Hive works on top the Hadoop HDSFS, it benefits from extraordinary feature of HDFS including Fault Tolerance, Reliability, High Availability, Scalability, etc. (Zhou, W., Feng, D., Tan, Z., & Zheng, Y., 2017). In addition, Hive can take advantage of distributed computing power of the cluster through assigning jobs to MapReduce, Tez and Spark engines to run complex queries. The study is focussed on studying of Hive Data Model and analysis of processing performance done by MapReduce and Tez (Hortonworks Inc., 2017).

Hive is open-source software that is components of Hadoop Ecosystem. It designed to query and analysis of huge amounts of data stored in HDFS using SQL-like language HiveQL (Hive Query Language). Hive also can be considered as ETL and Data warehousing tool for the large-scale data.

Hive can be also considered as alternative of the MapReduce with higher level of abstraction. Since MapReduce applications are developed in Java or Python, its more flexible, efficient and faster. It is designed to process structured data, so it suppose creation of table with certain structure before loading data. To work with Hive there are at least two options: Web GUI or more popular command line interface (CLI) using HQL (for DDL, less for DML). Hive supports four file formats those are TEXTFILE, SEQUENCEFILE, ORC and RCFILE (Record Columnar File).

Significant difference between DBMS and Hive is that: DBMS generally works on "Schema on READ and Schema on Write", but Hive on "Schema on READ only" (latest version on "WRITE Once READ Many Times").

## LITERATURE REVIEW

As the most popular platform for large-scale data management and analytics, Hadoop ecosystem has attracted substantial interest from researchers. Hadoop and its multiple components build unique system that allow to hide most of complexities staying focussed on real data analysis. This advantages inspires many researchers and us to to understand key components of the ecosystem in depth. The following studies from authors are devoted to different use-cases of the Hadoop Ecosystem.

Lee, Shao and Kang solved the problem of handling big graph that doesn't fit into memory of traditional system. Authors offered to use distributed platform Hadoop HDFS to store data physically and HBase for low latency data access (Lee H., Shao B., & Kang U., 2015). HBase is considered as an open source implementation of Google's Bigtable technology.

Authors of following paper explain relation between cloud computing platforms widely used by enterprises and Hadoop. Paper comprises the ways how enterprises can benefit from Hadoop platform along with existing cloud computing systems. Author observe and discuss primary sub-components of core Hadoop and how they related to each other enabling execution and monitoring of jobs that process data stored on top of HDFS (Ghazi M. R., & Gangodkar D., 2015).

Gadiraju, Verma, Davis and Talaga performed benchmark research comparing performance of Apache Hive with traditional database management system MySQL. Hive is a data warehouse platform that is member of Hadoop ecosystem and works on top of distributed file system HDFS. Hive queries written in HiveQL (SQL-like language) are executed as a MapReduse jobs using cumulative power of distributed Hadoop cluster (Laboshin, L. U., Lukashin, A. A., & Zaborovsky, V. S., 2017). The authors also provide evidence based on experiments that Hive loads the large datasets much faster than MySQL, while it loses its advantage over MySQL when loading the smaller datasets (Gadiraju K.K., Verma M., Davis K.C., & Talaga P.G., 2016). Same is true for query execution as well: Hive is much faster when it comes to processing large amounts of data. Using similar arguments, our paper paper states that the architecture of Hadoop and most of its components is tuned to manage huge amounts of data, but not for random low-latency data access to small chunks of data.

## HIVE AS A DISTRIBUTED ETL PLATFORM

### Hive Arcitectire

Even Hive provides same DDL and DML services and acts like DBMS, but in reality it is not a DBMS. Traditional DBMS is a software that encapsulates two main subcomponents implemented internally: storage and query engine. Unlike DBMS, Hive is just SQL Query Engine. Hive doesn't care about data storage, instead it relies on the scalable and redundant HDFS. Furthermore, Hive doesn't process computation-intensive queries itself, instead it consumes the MapReduce (or Tez and Spark) framework to use the distributed power of the Hadoop cluster. General architecture and interaction between it's sub-components is demonstrated on Figure 1.

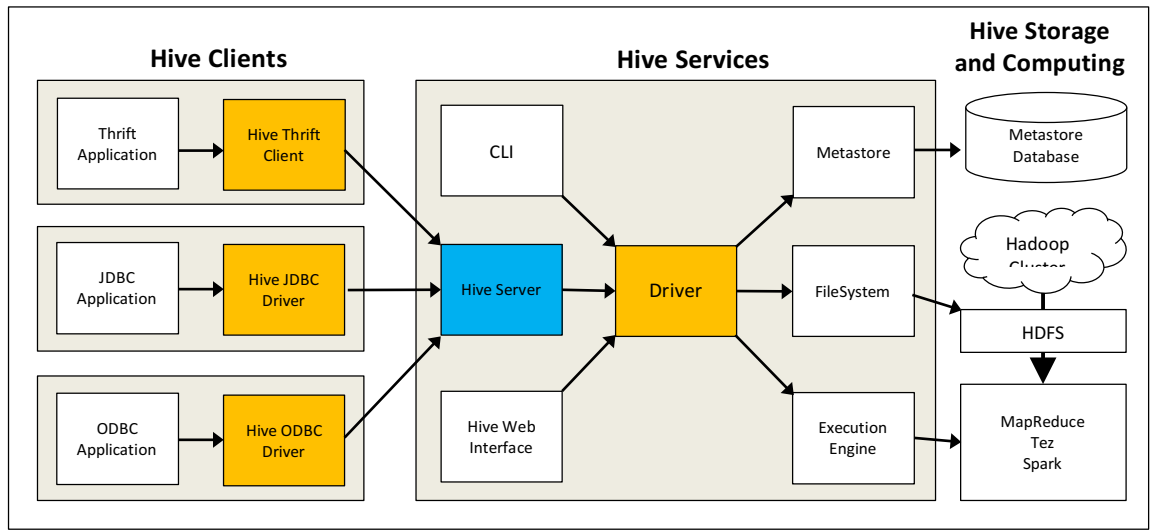

Figure 1: General Architecture of Hive.

**Hive Data Model – Schema on Read**

Unlike Database Systems, Hive enforces the Read schema rather than the Write schema. Any DBMS is strictly checking the model of any data that pretended to be inserted into database whether it follows to the predefined structure, and declines insertion if does not. Opposite to this Hive does not check the model of new data, but instead just copies it into HDFS without any control in order to improve writing speed. Hive checks the relevance of the data and the structure just on read.

Look at the following example that demonstrate what can happen if the data uploaded into table does not follow to the structure defined during the table creation (HiveQL script on Figure 2.).

```
create table salaries (id INT,
rank STRING,
discipline STRING,
yrsphd INT,
yrsservice INT,
sex STRING,
salary DOUBLE) row format delimited fields terminated by ',' stored as textfile
tblproperties("skip.header.line.count"="1");
```

Figure 2: Source code of HiveQL to create a table.

```
hive> describe salaries;
OK
id                      int
rank                    string
discipline              string
yrsphd                  int
yrsservice              int
sex                     string
salary                  double
Time taken: 0.814 seconds, Fetched: 7 row(s)
```

```
hive> select * from salaries limit 4;
OK
NULL    "Prof" "B"    19    18    "Male" 139750.0
NULL    "Prof" "B"    20    16    "Male" 173200.0
NULL    "Prof" "B"    30    23    "Male" 175000.0
NULL    "Prof" "B"    18    18    "Female"    129000.0
Time taken: 1.043 seconds, Fetched: 4 row(s)
```

```
[hadoop@namenode ~]$ hdfs dfs -cat
/user/hive/warehouse/mydb.db/salaries/Salaries.csv | head -n 5
"","rank","discipline","yrs.since.phd","yrs.service","sex","salary"
"1","Prof","B",19,18,"Male",139750
"2","Prof","B",20,16,"Male",173200
"3","Prof","B",30,23,"Male",175000
"4","Prof","B",18,18,"Female",129000
```

Figure 3: Source code of HiveQL to create a table.

As it has been clearly seen in Figure 3. the column "id" was declared as an integer. In this particular example, even data associated with mentioned column was loaded into Hive, but still the values are missing in output of the "select" query. The problem becomes clear after screening the content of data source file. The values of column associated with attribute "id" is surrounded by quotes, what means the type is character. The data type conflict has not been revealed during data load, but comes clear on read.

**Hive Metadata**

The first implementation of Metadata in Hadoop Ecosystem has started with Hive that used Metastore to store description of data model of Hive tables. Later it became clear that other components of Hadoop need this technique as well. As a result, new components, particularly HCatalog and WebHCat (REST API) appeared those enable Hive metastore to other components of the Hadoop Ecosystem.

Hive storage consists of two categories: metadata and real data. Metadata of tables is stored in "Meta storage database" (generally MySQL), while real data stored in HDFS.

**Data Model Components**

The Hive data models contain the following components:
- Databases

- Tables
- Partitions
- Denormalizing
- Buckets or clusters

Data partitioning is about splitting datasets into smaller pieces in order to avoid reading huge volumes of data at once. The reason of doing this is to reduce the speed of read and manipulation of any particular data. Unlike traditional DBMS systems, HDFS does not support low latency transactions, instead it was designed to support the ingestion of huge amounts of data at high speed. In terms of Hive, Data Partitioning enables breaking the data into smaller subsets that allow to retrieve particular data enclosed into subset instead of retrieving all data from the table. (Gwen Shapira, Jonathan Seidman, Ted Malaska, Mark Grover, 2015).

To create a partitioned table in Hive, the certain instruction of HiveQL "PARTITIONED BY" should be used while creating the table (as shown in Figure 4.).

```
CREATE TABLE customer (id INT, name STRING, surname STRING) PARTITIONED BY (city STRING);

LOAD DATA LOCAL INPATH '/data/customer.txt' INTO TABLE customer PARTITION (city STRING);
```

Figure 4: HiveQL code to Create and Manage Partitioned Table.

After loading the data into partitioned table, the content of folder on HDFS associated with the table may look similar to the structure demonstrated on Figure 5.

```
/apps/hive/warehouse/customer
|---Baku/
|    |---file1
|    |---file2
|    |---file3
|
|---Sheki/
|    |---file1
|    |---file2
|
```

Figure 5: The Structure of Partitioned Table's Folder.

HiveQL provides special command (SHOW PARTITIONS table;) to list all partitions generated for the table.

Another technique that can increase data access and processing speed on expense of storage usage effectiveness is Data Denormalization. Usually, tables in relational databases follow to the requirements of the 3rd Normal Form (3NF), as well as 1st and 2nd. The 3NF states that if two tables are related between each other based on Primary Key (example on Figure 6.), not any attribute of the master table can be included into the second table except Primary Key itself. This approach helps to keep records smaller saving a memory at the same time providing high consistency of data.

```
ID   | Name  | Surname   | City   |
-----|-------|-----------|--------|
1001 | Ali   | Kerimov   | Baku   |
1002 | Samir | Alasgarov | Sheki  |
1003 | Jemil | Mamadov   | Baku   |

FID  | Product  | Qty | Date       |
-----|----------|-----|------------|
1001 | HDD      | 2   | 10.12.2017 |
1001 | Keyboard | 1   | 10.12.2017 |
1002 | CPU      | 1   | 10.12.2017 |
1003 | Printer  | 1   | 10.12.2017 |
```

Figure 6: Two Related Table those follow to 3rd Normal Form.

When it comes to Hadoop, in reality each join-included query is accomplished by the MapReduce operation that takes too much resources of cluster, especially if the query is called frequently. The idea is to change the structure of table in advance that will eliminate the need to join tables on run as it is shown in Figure 7.

```
FID  | Name  | Surname   | City  | Product  | Qty | Date       |
-----|-------|-----------|-------|----------|-----|------------|
1001 | Ali   | Kerimov   | Baku  | HDD      | 2   | 10.12.2017 |
1001 | Ali   | Kerimov   | Baku  | Keyboard | 1   | 10.12.2017 |
1002 | Samir | Alasgarov | Sheki | CPU      | 1   | 10.12.2017 |
1003 | Jemil | Mamadov   | Baku  | Printer  | 1   | 10.12.2017 |
```

Figure 7: Same two table are Denormalized by joining in advance.

## DISTRIBUTED QUERY PROCESSING

HiveServer2 (HS2) is essential component of Hive 2.x that in contrast with HiveServer1, supports multiclient concurrency and authentication. Another important advantage of the HiveServer2 is the fact that it provides JDBC and ODBC interface to interact with Hive.

Queries submitted to Hive are processed in the following way (look at Figure 8.):

1. Client send query to one of HiveServer2 instances connecting over JDBC/ODBC interface;
2. Query is compiled and divided into sub-tasks by HS2;
3. Compiled query is submitted to Tez or MapReduce (depending on which execution engine was set);
4. Coordinator (Tez/ApplicationMaster) asks YARN for allocation of the computing resources (containers) across the cluster;
5. Tez/ApplicationMaster transfers tasks into containers;
6. Data that resides within HDFS in variety formats (text, ORC, AVRO, Parquet) is read using HDFS interface;
7. Data is processed and result are returned over JDBC/ODBC interface.

Tez is new high-performance batch processing framework for execution of complex Hive queries that significantly outperforms traditional MapReduce framework (which is used by Hive as a default execution engine).

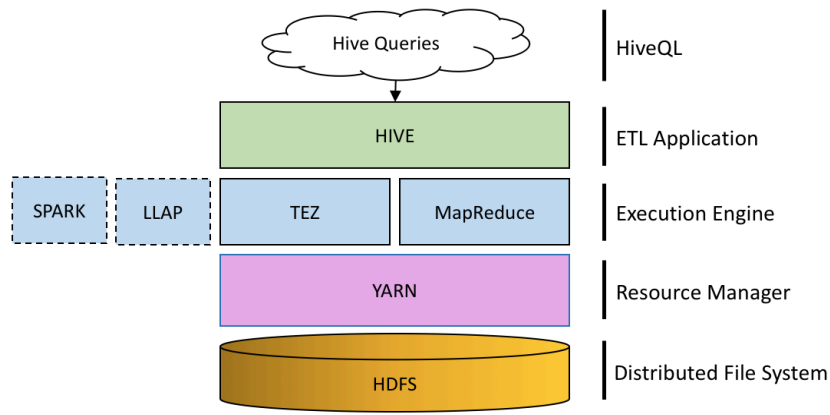

Figure 8: Query Execution Architecture of Hive.

Hive queries are submitted to HiveServer2 server that generates Tez graph that in its turn is transfered to YARN for processing. Each Hive query is monitored by their individual Tez ApplicatiuonMaster. Number of simultaneous queries are limited with number of allowed ApplicationMasters.

MapReduce and Tez have significant differences in computation models that effects their performance. Looking to the architecture of computation model of MapReduce shown in Figure 10., (while executing the query shown on Figure 9.) it obvious that following elements increase cost and time of the MapReduce execution:
- To execute this query using MapReduce execution engine, Hive should launch 4 MR jobs
- Generally, each MR job has its own start-up time and after processing writes result to HDFS providing data to subsequent job for read.

```
SELECT a.occ_code, c.occ_name, COUNT(*) AS cnt, AVG(b.value) AS avg
FROM occup a
JOIN occupdata b ON (a.sid = b.sid)
JOIN jobs c ON (a.occ_code = c.occ_code)
GROUP BY a.occ_code, c.occ_name
ORDER BY avg DESC
```

Figure 9: HiveQL code to retrieve data from three tables.

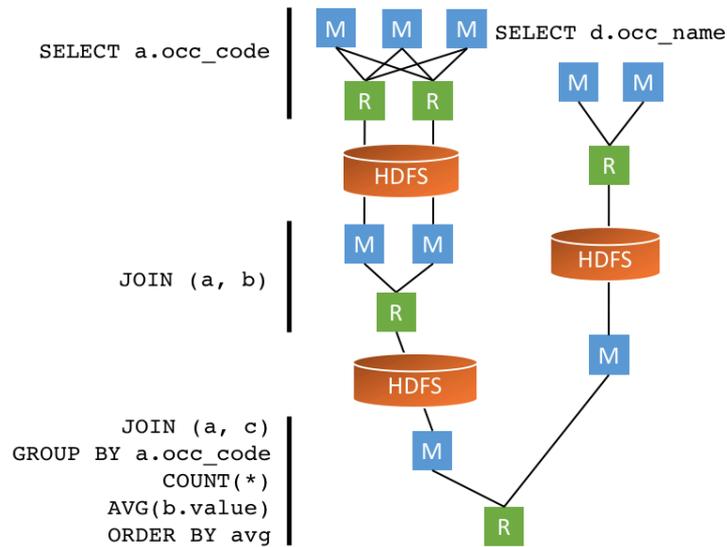

Figure 10: Computation Model of MapReduce.

At the same the following features of Tez make it's computation model (look at Figure 11.) more efficient and consequently fast:
- In contrary to MapReduce, Tez performs complex query as a single execution graph
- Tez doesn't implement wasting intermediate IO operations with HDFS
- Vertexes in graph are processing jobs and edges are data streams
- Tez supports "hot containers" to start jobs immediately without wasting time for start-up

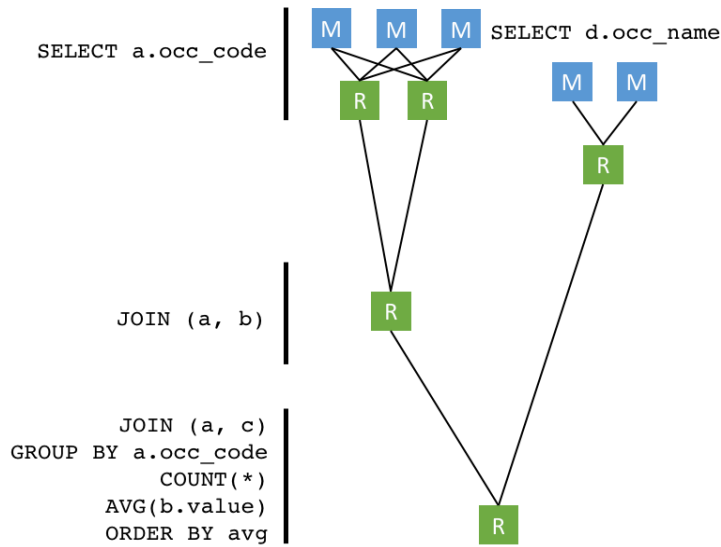

Figure 11: Computation Model of Tez.

LLAP is a new computation paradigm recently implemented in Hive. It consists of the set of persistent daemons that execute fragments of Hive queries. This persistency allows to start jobs much faster, since containers do not need warm-up. Query execution on LLAP is very similar to Hive without LLAP, except that jobs run inside LLAP daemons, and not within YARN containers. Both the Hive on Tez engine for batch queries and the enhanced Hive on Tez LLAP-enhanced engine run on YARN nodes. The Hive LLAP layer over Tez execution engine requires particular Hadoop YARN settings to consume full potential of this new advancement in Hive (Hortonworks Community Connection, 2017).

**Computing Resources for Experiments**

In the framework of the Center for Data Analytics Research (CeDAR) the mid power computing cluster has been launched (Figure 12.).

**Computing Cluster Hardware** – the primary component of the CeDAR. This is powerful, scalable and fault-tolerant computing cluster based on distributed architecture, which operates totally on open-source software. Each computing node is equipped with Intel Xeon E-5 processor, 96 GB memory, 8 LFF Hard Drives of 2 TB storage each and 1Gb Ethernet support.

Characteristics of the cluster:
- Processing Cores: 102
- RAM: 1,568 TB
- Storage: 136 TB

Specifications of the cluster's components:

- **NameNode (1 unit):** HP DL360 Gen9 4LFF CTO Server: 2 x Intel® Xeon® E5-2603v4 (1.7GHz/6- core/15MB/85W), 128GB RAM, 4 x HP 2TB 12G SAS 7.2K rpm LFF HDD, HP 1U LFF Gen9 Mod Easy Install Rail Kit;
- **DataNode (15 units):** HP DL380 Gen9 12LFF CTO Server: Intel® Xeon® E5-2603v4 (1.7GHz/6- core/15MB/85W), 96GB RAM, 8 x HP 2TB 12G SAS 7.2K rpm LFF HDD, HP 2U LFF Easy Install Rail Kit;

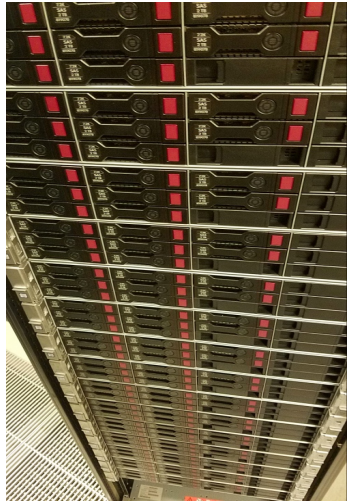

Figure 12: Distributed cluster at the CeDAR research center.

**Computing Cluster Software** – the cluster is running on Apache Hadoop ecosystem. It is deployed using Hortonworks HDP 2.6 distribution, which is 100% open source.

**EXPERIMENTAL RESULTS**

Many experiments have been accomplished on the CeDAR cluster to identify most important parameters and criteria those have highest impact on the performance of the query processing on Hive.

The following datasets listed in Table 1. publicly available at the Kaggle Datasets repository (Kaggle Inc., 2018) were used to implement experiments. The table "houses_part" with relatively large number of records was used to reveal the effect of the file format on the performance. Other three tables "occup", "occupdata" and "jobs" were used in the complex HQL queries where 2-4 tables are merged using multiple join instructions.

Table 1: Parameters of tables used in experiments.

|    | Name        | Num. of Records | Num. of Fields | File size |
|----|-------------|-----------------|----------------|-----------|
| 1. | houses_part | 22.489.349      | 10             | 2.4 GB    |
| 2. | occup       | 6.462.646       | 15             | 1.4 GB    |
| 3. | occupdata   | 6.462.646       | 5              | 417 MB    |
| 4. | jobs        | 1.090           | 5              | 0.05 MB   |

**Query Performance depending on Execution Engine**

As it was stated above Tez outperforms MapReduce as a execution engine while processing HiveQL queries. Computation architectures of both engines depicted on Figures 10. and 11. displays key differences those affect the performance. As the Table 2. demonstrates, based on experimental query execution initiated with three different queries it is clearly seen that Tez is faster.

Table 2: Query performance dependence on Execution Engine.

|   | Query | Tez (sec.) | MapReduce (sec.) |
|---|---|---|---|
| 1. | select propertytype, count(*) from houses group by propertytype; | 25 | 34[*] |
| 2. | select PropertyType, sum(price) as sumprice, count(*) from houses group by PropertyType; | 28 | 33[*] |
| 3. | set hive.input.format=org.apache.hadoop.hive.ql.io.HiveInputFormat; set hive.merge.mapfiles=false; select PropertyType, sum(price) as sumprice, count(*) from houses group by PropertyType; | 32 | 154[**] |

[*] - with following settings
```
set hive.input.format=org.apache.hadoop.hive.ql.io.CombineHiveInputFormat;
set hive.merge.mapfiles=true;
```

[**] - if execution engine is tuned using following settings, the execution time increases to about 5 times (154 sec):
```
set hive.input.format=org.apache.hadoop.hive.ql.io.HiveInputFormat;
set hive.merge.mapfiles=false;
```

**Partitioning Effect on Query Performance**

As it was indicated above, Hive supports the partitioning of the data file by value of specific column or several columns. This technique can significantly effect the query performance. This impact can be explained by the fact that in the partitioned table the query executor does not forced to read whole file from the HDFS, instead it reads just particular partition(s) where the data of interest is located.
To create the partitioned table, the instruction PARTITIONED BY "columnName TYPE" should be included into the table creation script (look at Figure 13.). To ingest the data into the partitioned table, we need to load data firstly into non-partitioned intermediate table, and after that insert into the partitioned table.

```
CREATE TABLE houses_part (
id STRING, price STRING, dateoftransfer STRING, oldNew STRING, duration
STRING, city STRING, district STRING, county STRING, ppd STRING, status
STRING)
PARTITIONED BY (propertytype STRING)
ROW FORMAT DELIMITED FIELDS TERMINATED BY ',' STORED AS TEXTFILE
TBLPROPERTIES ("skip.header.line.count"="1");

INSERT INTO TABLE houses_part
PARTITION (propertytype)
SELECT id, price, dateoftransfer, oldNew, duration, city, district, county,
ppd, status, propertytype
FROM houses;
```

Figure 13: Storage Efficiency by Hive File Formats.

After ingesting the data into table, the exact number of data-files equal to the number of unique values of column. These files (partitions) will appear in the folder associated with partitioned table. In presented example (Figure 13.) there are five distinct values ("D", "F", "O", "S", "T") of the column and accordingly there are five files (the size of each partition is on the left) in the directory.

```
[root@nnode ~]# hdfs dfs -du /apps/hive/warehouse/bigdata.db/houses_part
543805800   /apps/hive/warehouse/bigdata.db/houses_part/propertytype=D
427515342   /apps/hive/warehouse/bigdata.db/houses_part/propertytype=F
10602972    /apps/hive/warehouse/bigdata.db/houses_part/propertytype=O
653235054   /apps/hive/warehouse/bigdata.db/houses_part/propertytype=S
725547877   /apps/hive/warehouse/bigdata.db/houses_part/propertytype=T
```

Figure 13: Computation Model of Tez.

The experimental results obtained after execution the same queries on the data that this time stored in partitioned table reveals significant performance improvement, as shown on Table 3., in comparison with results of unpartitioned table shown in Table 2.

Table 3: Folder associated with partitioned table.

|    | Query | Tez (sec.) |
|----|-------|------------|
| 1. | select PropertyType, count(*) as count from houses_part group by PropertyType; | 5.5 |
| 2. | select PropertyType, sum(price) as sumprice, count(*) from houses_part group by PropertyType; | 13.2 |

**File Format Effect on Query Performance**

Hive as a many other components of the Hadoop Ecosystem is designed following to the write-once concept, but not for low latency data access. By default, Hive that is functioning on top of HDFS supports neither ACID nor OLTP transactions. Even so, low latency data access can be enabled using particular file format to store the data and LLAP (Low Latency Analytical Processing) daemons based on persistent query executors.

Hive supports four file formats those are TEXTFILE, SEQUENCEFILE, ORC and RCFILE (Record Columnar File). Optimized Row Columnar (ORC) is a file format that specially designed for storing the Hive data. ORC outperforms all other file formats supported by Hive. The following chart in Figure 14. demonstrates the advantage of the ORC file format in terms of storage efficiency.

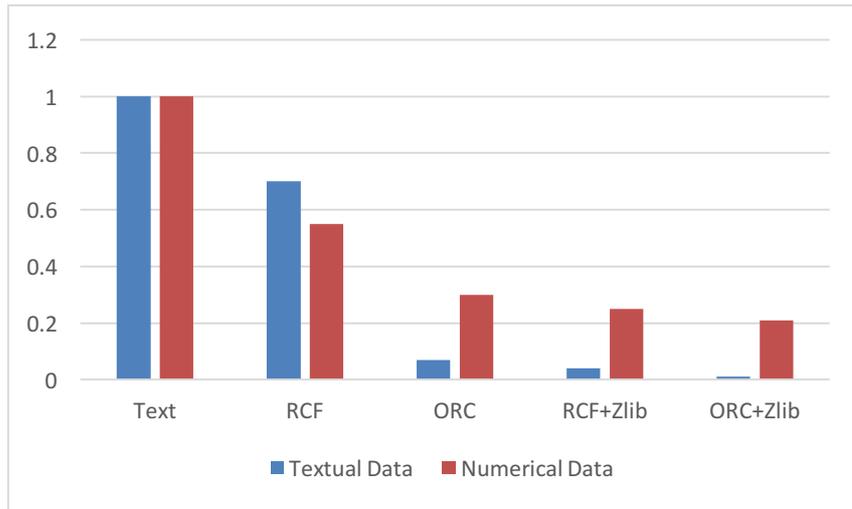

Figure 14: Storage Efficiency by Hive File Formats.

Besides, ORC file format supports internal indexing that enables skipping large intervals of rows those are out of interest. Default size of the ORC blocks is 256 MB what makes sequential read highly effective and decrease the load on the NameNode (Blog by Christian Prokopp, 2014).

In order to create the table that stores data in ORC format, it is enough to replace the instruction "STORED AS TEXTFILE" to "STORED AS ORC" in HiveQL code shown in Figure 13.

After loading the data into ORC-table, the following files will be created in the folder associated with partitioned table. Simple comparison of total size of the folders associated with two tables stored as TextFile and ORC, demonstrates that ORC file format is about 4 times more space-effective even without compression (look at Figure 15.).

```
[root@nnode ~]# hdfs dfs -du /apps/hive/warehouse/bigdata.db/houses_part_orc
138576786   /apps/hive/warehouse/bigdata.db/houses_part_orc/propertytype=D
106081995   /apps/hive/warehouse/bigdata.db/houses_part_orc/propertytype=F
1016862     /apps/hive/warehouse/bigdata.db/houses_part_orc/propertytype=O
166501388   /apps/hive/warehouse/bigdata.db/houses_part_orc/propertytype=S
183961469   /apps/hive/warehouse/bigdata.db/houses_part_orc/propertytype=T

[root@nnode ~]# hdfs dfs -du /apps/hive/warehouse/bigdata.db/
2405685902  /apps/hive/warehouse/bigdata.db/houses
2360707045  /apps/hive/warehouse/bigdata.db/houses_part
596138500   /apps/hive/warehouse/bigdata.db/houses_part_orc
```

Figure 15: Folder associated with partitioned table stored as ORC file.

By being space-effective ORC file format has quite significant impact on overall performance of the HiveQL execution. If same data occupies less space on the file system, it means less time will be spent for HDFS read operations speeding up split and map jobs. The Table 4. shows the time spent for query execution applied to the table stored as ORC file. It is important to notice that the execution performance will increase in parallel with growth of data size.

Table 4: Query Performance on the ORC file format.

|   | Query | Tez (sec.) |
|---|---|---|
| 1. | select PropertyType, count(*) as count from houses_part_orc group by PropertyType; | 4.8 |
| 2. | select PropertyType, sum(price) as sumprice, count(*) from houses_part_orc group by PropertyType; | 11.6 |

## CONCLUSION AND FUTURE WORK

The Hadoop Ecosystem becomes the de facto standard platform of choice for enterprises that provides critical features like scalability, fault-tolerance, low TCO and high ROI those are hardly available in traditional IT platforms. One of the most important components of Hadoop Ecosystem Apaches Hive has been thoroughly observed, key features those have highest impact on performance revealed and extensive experiments conducted to demonstrate the truth of findings.

While observing the results it is important to keep in mind that Hive, as a many other components of Hadoop Ecosystem running on top of HDFS, is not designed for low latency random access to data. Real power of Hive can be seen while processing huge chunks of structured data stored on HDFS.

Further research is needed to investigate the computation model of LLAP and its advantage in terms of the query processing performance as compared to Tez without LLAP. Even both Tez and Tez with LLAP are working on top on the YARN nodes, there are some peculiarities implemented in the architecture of LLAP engine those make the query processing about 25 times faster than performance offered by Hive without LLAP (Nita Dembla, 2016).

## ACKNOWLEDGEMENT

This research was supported by a grant from "Strengthening Teaching and Research Capacity at ADA University" project funded by the European Union. For the research experiments the distributed computing cluster of the Center for Data Analytics Research (CeDAR) has been used.